\documentclass[conference,10pt]{IEEEtran}

\usepackage{graphicx}
\usepackage{tikz}
\usepackage{siunitx}
\usepackage{enumitem}
\usepackage{float}
\usepackage{threeparttable}
\usepackage{booktabs}
\usepackage{subcaption}
\usepackage{xcolor}
\usepackage{colortbl}
\usepackage{amsmath}
\usepackage{amssymb}
\usepackage{pifont}

\newcommand{\cmark}{\ding{51}}
\newcommand{\xmark}{\ding{55}}

\usepackage{adjustbox}
\usepackage{cite}
\usepackage{url}
\usepackage{balance}   

\usepackage{geometry}
\newcommand{\fakepara}[1]{%
  \addvspace{0.4em}%
  \par\noindent\textbf{\textit{#1:}}%
}

\definecolor{darkblue}{RGB}{0,51,102}

\newcommand{\circnumblue}[1]{%
  \tikz[baseline=(char.base)]\node[shape=circle,fill=darkblue,inner sep=1pt,
  text=white,font=\small] (char) {#1};}

\newcommand{\circnumblack}[1]{%
  \tikz[baseline=(char.base)]\node[shape=circle,fill=black,inner sep=1pt,
  text=white,font=\small] (char) {#1};}

\newcommand{\circG}{%
  \tikz[baseline=(char.base)]\node[shape=circle,fill=black,inner sep=2pt,
  text=white,font=\bfseries\small] (char) {G};}

\newcommand{\dg}[1]{\textbf{[D#1]}}

\begin{document}

\title{Motion-Coupled Sensing:\\
When the State Change Powers Its Own Sensing}

\author{%
  \IEEEauthorblockN{%
    Muhammad Tahir,\;
    Muhammad Mubbashar Baig,\;
    Umer Irfan,\;
    Muhammad Ahad,\;
    Naveed Anwar Bhatti}
  \IEEEauthorblockA{%
    Department of Computer Science,
    Lahore University of Management Sciences (LUMS), Lahore, Pakistan\\
    \{24030017, baig.muhammad, 27100363, 22260007, naveed.bhatti\}@lums.edu.pk}}

\maketitle

\begin{abstract}

Batteryless IoT systems have largely followed two paths: ambient-energy
sensing, where energy arrival is decoupled from the event being monitored,
and kinetic event telegrams, where a user actuation powers a short report
of the actuation itself. Mechanically gated states expose a third case: the
access motion is not only an event to report, but the moment at which a
latent physical state may have changed and must be measured. We show that
routine hinge motion can supply enough energy for one bounded
wake--sense--transmit transaction, including ultrasonic sensing and a long-range
LoRa uplink. We call this principle \emph{motion-coupled sensing} and
instantiate it with an open-source compact electromagnetic harvester
that retrofits to bins, doors, and cabinets with no structural
modification. We size the platform for the most demanding workload, waste-bin
monitoring, where each actuation must power both an ultrasonic measurement and a long-range LoRa uplink. Across five campus locations and
5{,}945 lid actuations, the bin deployment achieves \textbf{99.3\%
per-event transmission reliability}. Field deployments on room doors with 1{,}870 actuations
and office cabinets with 1{,}636 actuations achieve
\textbf{92\%} and \textbf{94\%} transmission success respectively,
demonstrating that the same energy envelope transfers across hinge
geometries without hardware redesign. These results show that
mechanical access can be treated as a self-powered sensing transaction,
removing periodic polling and scheduled battery maintenance for IoT deployments.
\end{abstract}

\section{Introduction}
\label{sec:introduction}

Many IoT deployments monitor physical states that can change only
through a mechanical action. A waste bin's fill level changes
when its lid opens; a room is entered when its door swings; a cabinet is accessed when its hinge moves. For such mechanically gated
states, battery-powered periodic polling is structurally inefficient:
the device spends energy checking for changes that are physically
impossible between access events. This inefficiency compounds
the operational burden of replacement schedules, silent node failures,
and battery disposal~\cite{ref38,ref41}. Solar harvesting can reduce
this burden outdoors, but it remains unreliable for sensors installed
indoors, in shaded areas, or inside enclosed objects~\cite{ref8}.

This paper treats mechanical access not as a nuisance to be monitored,
but as an energy source to be exploited. The same action that changes
the monitored state releases a short burst of kinetic energy. If that
energy can be captured and conditioned, it can power the full
wake--sense--transmit pipeline: the node harvests energy from the access, 
wakes its electronics, senses or encodes the event, transmits a
packet, and returns to zero idle current. We call this design
pattern \textbf{motion-coupled sensing}: for mechanical-access objects,
the motion that creates the information also triggers and powers the
sensing transaction.

Conventional IoT nodes are
optimized for long lifetime, duty cycling, or periodic sampling. A
motion-coupled node is instead optimized for \emph{per-event
sufficiency}: each access event must provide enough usable energy to
complete one bounded transaction. This removes scheduled battery
maintenance and eliminates idle draw by leaving active electronics
unpowered between access events.

\begin{figure*}[t]
\centering
\includegraphics[width=0.8\linewidth]{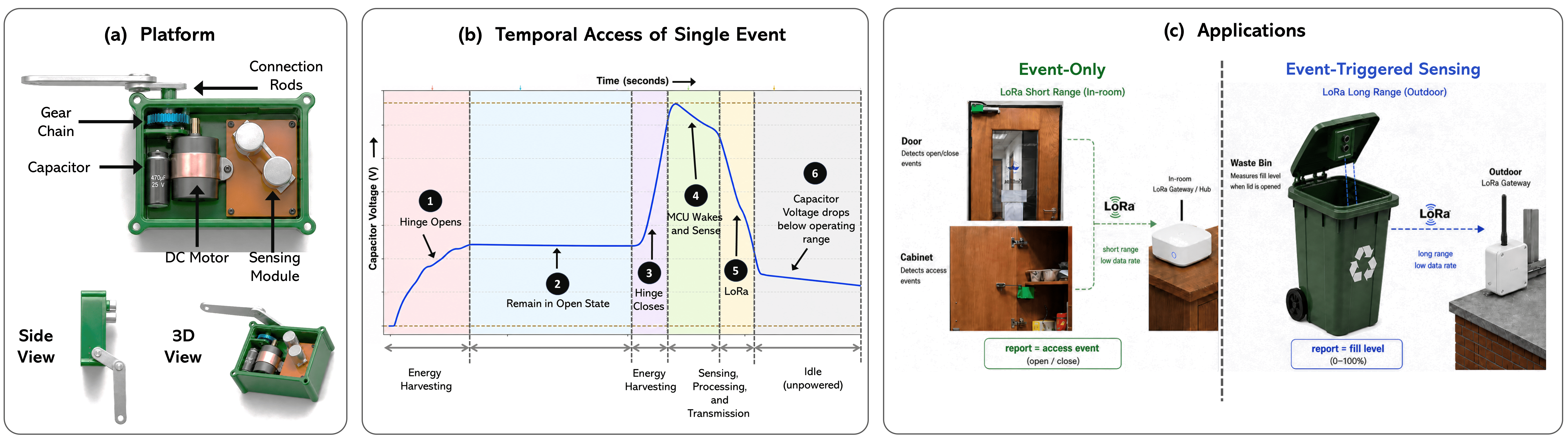}
\caption{Motion-coupled sensing as an atomic access transaction.
\textbf{(a)}~Platform retrofits to hinges such as bins, doors,
and cabinets. \textbf{(b)}~Access motion is not only an energy source; it
also defines the state-validity boundary, wake-up trigger, and power source
for the complete sense-and-send cycle. \textbf{(c)}~Same hardware
supports two subclasses: \emph{event-only} reporting and \emph{event-triggered sensing}.}
\label{fig:overview}
\end{figure*}

Motion-coupled applications fall into two subclasses
(Fig.~\ref{fig:overview}c). In \emph{event-only} applications, the
access event itself is the information of interest; for example, a door
or cabinet opening only needs to be encoded and reported. In
\emph{event-triggered sensing} applications, the access event supplies
the trigger and energy, but the node must also measure a state affected
by that access. Waste-bin monitoring is the canonical example: opening
the lid is when the fill state may change, but the node must still
measure the resulting fill level before transmission. This makes the bin
variant the highest-energy workload because each actuation must power
both active sensing and a long-range wireless uplink. The application is
also practically important: poor solid waste management imposes
significant public health burdens in developing regions~\cite{ref4,ref2}.

We present an
open-source\footnote{Hardware designs, firmware, and analysis scripts
are available at:
\url{https://github.com/SYSNET-LUMS/Batteryless-event-driven-sensing-platform.git}}
batteryless platform built around a compact 
electromagnetic harvester that retrofits onto hinged access points
without structural modification (Fig.~\ref{fig:overview}a). A single actuation charges a capacitor, wakes a low-power MCU,
performs sensing or event encoding, and transmits one LoRa packet. The
active transaction completes in under 0.5\,s, after which the node
returns to zero idle current (Phases \circnumblack{4} and
\circnumblack{5} in Fig.~\ref{fig:overview}b). The same LoRa transceiver
family supports both deployment regimes: spreading factor 10 (SF10) for
outdoor long-range uplinks, and spreading factor 6 (SF6) for indoor short-range uplinks to nearby hubs.

The platform is sized based on measured access behavior. We instrumented waste bins with a rotary encoder and collected 832 lid interactions across five locations, measuring angular displacement, timing, and hinge-speed range. These measurements guided the generator selection and a 1:42.6 gear-train ratio, chosen to support the most demanding transaction: ultrasonic fill-level sensing plus a long-range LoRa uplink from one lid actuation. We evaluated the same platform across three hinged-access deployments. In outdoor waste-bin deployments, the platform achieves \textbf{99.3\% per-event transmission reliability} over 5{,}945 lid actuations in three weeks. In indoor deployments, the same core achieves \textbf{92\%} success over 1{,}870 room-door actuations and \textbf{94\%} success over 1{,}636 cabinet actuations. These results show that a harvester sized for the highest-energy sensing workload transfers to lower-energy event-only monitoring across different hinge geometries at deployment scale.

\noindent\textbf{Contributions.}
\begin{enumerate}[leftmargin=2.5em,label=\textbf{[C\arabic*]}]

\item \textbf{Mechanically gated sensing as a self-powered transaction.}
We identify motion-coupled sensing as a class of IoT workloads where
mechanical access defines the state-validity boundary, supplies the energy,
triggers the node, and defines the reliability unit. Our event-triggered
workload uses one access to power active sensing plus long-range LoRa
reporting of the resulting state.

\item \textbf{Transaction-sized harvester design.}
We derive a reproducible batteryless harvester from 832 measured lid
actuations across five campus sites, translating a 17.85--27.8\,RPM
hinge-speed envelope into a 1:42.6 gear train, 1000\,$\mu$F buffer, and
11.5\,V wake-up threshold for a measured $\sim$61\,mJ
wake--sense--transmit transaction.

\item \textbf{Field validation on the highest-energy workload.}
We validate the platform on outdoor waste-bin sensing, the most
demanding workload, across five campus locations and 5{,}945 lid
actuations over three weeks, achieving \textbf{99.3\% per-event
transmission reliability}.

\item \textbf{Multi-workload field validation across access types.}
We field-deploy the same platform on room doors with 1{,}870 actuations and office cabinets
with 1{,}636 actuations, achieving \textbf{92\%} and
\textbf{94\%} per-event transmission success respectively without
hardware redesign, demonstrating that the bin-sized energy envelope
transfers across hinge geometries at deployment scale.

\end{enumerate}

\section{Related Work}

\subsection{Smart Environments and Access Monitoring}

Smart-bin systems commonly pair ultrasonic or infrared sensing with
cellular, Wi-Fi, or LPWAN links to report fill level and schedule
collections~\cite{ref5,ref21}. Commercial systems extend lifetime with
large batteries or solar/hybrid supply~\cite{ref7,ref10}, while
solar LoRaWAN smart-bin nodes show that batteryless waste monitoring is
possible when outdoor light is available~\cite{ref8}. These designs
reduce collection overhead, but they still treat sensing energy and bin
access as separate concerns: the node is powered from a battery, solar
source, or ambient harvester rather than from the lid motion that makes
fill-level change possible.

Access monitoring in smart homes and buildings uses PIR sensors, reed
switches, and BLE tags to detect occupancy or door/window state
changes~\cite{Suryadevara2012HomeMonitoring,Kelly2013IoTHomeDoor}.
Commercial kinetic building controls further show that a short human
actuation can power a maintenance-free radio telegram~\cite{EnOcean2020World}.
Kinetic switches are the event-only endpoint of this design space: the
actuation powers a short telegram reporting the actuation itself.
Motion-coupled sensing targets the harder event-triggered case, where the
access motion powers a post-access measurement of the physical state whose
validity is bounded by that same access.

\subsection{Batteryless Communication and Intermittent Sensing}

Battery-free RFID and backscatter platforms such as WISP and ambient
backscatter demonstrate sensing and communication without batteries by
harvesting or reflecting RF energy~\cite{ref12,ref13}. Intermittent
computing systems such as Mementos and Mayfly provide software support
for long-running sensing tasks across repeated power failures~\cite{ref14,Hester2017Mayfly}. Batteryless LoRaWAN work has also modeled how
capacitor-backed nodes should schedule sensing and transmission under
scarce harvested energy~\cite{Sabovic2020EnergyAware}. These systems
are important substrates for batteryless IoT, but they assume external
RF, light, or other ambient energy and do not use the state-changing
mechanical access event itself as the energy source, trigger, and
transaction boundary.

\subsection{Kinetic and Event-Driven Harvesting}

Kinetic energy harvesting has been explored with piezoelectric,
electromagnetic, and rotational transducers for self-powered IoT
systems~\cite{ref22,ref16}. Event-driven harvesting nodes
can recognize vibration events and transmit short alarms~\cite{Liu2023EventDriven}, and energy-neutral localization systems use
motion-triggered operation to reduce idle cost~\cite{Mayer2022UWB}.
These systems target ambient vibration, human motion, or localization
updates rather than mechanically gated object states. Motion-coupled
sensing is therefore not generic kinetic harvesting: it uses kinetic energy
only when that motion is causally tied to the state being monitored. In the
bin workload, the access does more than announce itself; it powers the
measurement that validates the resulting fill state. Table~\ref{tab:related_gap} summarizes the research gap this work is addressing.

\begin{table}[t]
\centering
\caption{Motion-coupled sensing relative to adjacent smart-sensing,
batteryless, and kinetic systems.}
\label{tab:related_gap}
\scriptsize
\setlength{\tabcolsep}{2.2pt}
\renewcommand{\arraystretch}{1.06}
\adjustbox{max width=\columnwidth}{%
\begin{tabular}{@{}p{0.37\columnwidth}p{0.19\columnwidth}p{0.15\columnwidth}p{0.10\columnwidth}p{0.14\columnwidth}@{}}
\toprule
\textbf{System class} &
\textbf{Energy model} &
\textbf{Post-access sensing?} &
\textbf{Range} &
\textbf{Energy tied to access?} \\
\midrule
Battery smart-bin/access~\cite{ref5,ref21,Suryadevara2012HomeMonitoring,Kelly2013IoTHomeDoor}
    & stored battery & \cmark & long & \xmark \\
Solar batteryless bin~\cite{ref7,ref10,ref8}
    & ambient light & \cmark & long & \xmark \\
Kinetic event telegram~\cite{EnOcean2020World}
    & actuation burst & \xmark & short & \cmark \\
RFID/backscatter~\cite{ref12,ref13}
    & reader/RF field & limited & short & \xmark \\
Intermittent batteryless~\cite{ref14,Hester2017Mayfly,Sabovic2020EnergyAware}
    & ambient & chunked & varies & \xmark \\
Event-driven kinetic~\cite{ref22,ref16,Liu2023EventDriven,Mayer2022UWB}
    & motion & limited & short & no \\
\textbf{This work}
    & \textbf{access-motion burst} & \cmark & \textbf{SF10 long} & \textbf{\cmark} \\
\bottomrule
\end{tabular}}
\vspace{-14pt}
\end{table}

\section{Platform Design}
\label{sec:design}

The platform implements the motion-coupled sensing transaction introduced in Section~\ref{sec:introduction}. Each access event must satisfy three invariants: the motion must create or reveal the information of interest, the same motion must provide the energy burst to power the node, and the computation must be bounded so that one stored energy burst is sufficient to complete a wake--sense--transmit cycle. These invariants shift the design problem from continuous energy availability to \emph{per-event sufficiency}. Unlike ambient harvesters, motion-coupled harvesting aligns energy arrival with moments when the monitored state can change. The design problem is therefore not long-term energy neutrality, but whether each access can provide enough energy for one bounded transaction. Rather than maintaining a node in a low-power polling state, the platform keeps active electronics unpowered until an access event has charged a local energy buffer above the threshold required for one reliable packet. Figure~\ref{fig:overview} summarizes the resulting architecture: a shared harvesting core captures hinge motion, while the sensing workload and radio configuration vary with the deployment.

\subsection{Design Goals}
\label{sec:design-goals}

Four goals guide the platform design.

\begin{enumerate}[leftmargin=2.5em,label=\textbf{[D\arabic*]},
                  topsep=2pt,itemsep=1pt]

\item \textbf{No scheduled battery maintenance:}
      Operate without primary or rechargeable batteries, eliminating
      replacement, recharging, and battery-aging failure modes.

\item \textbf{Retrofittable mechanical integration:}
      Attach to common hinged access points without permanent
      modification to the host object.

\item \textbf{Commodity, low-cost construction:}
      Use inexpensive and locally available parts, including DC motors
      as generators, spur gears, low-power microcontrollers, and
      off-the-shelf radios and sensors.

\item \textbf{Deployment-matched communication:}
      Use the same LoRa transceiver across deployments,
      with higher spreading factors for outdoor long-range bins and
      lower spreading factors for short-range indoor doors and cabinets.

\end{enumerate}

\subsection{Platform Architecture}
\label{sec:architecture}

The system has three logical tiers. \textbf{(1)} At the edge, a batteryless node harvests kinetic energy from the access event and uses the stored burst to transmit either an access-event packet or a sensor measurement. \textbf{(2)} At the communication tier, packets are received by either an outdoor LoRaWAN gateway for long-range deployments or a nearby in-room LoRa hub for short-range deployments. \textbf{(3)} At the backend, packets are stored and optionally converted into alerts or downstream actions.

The central architectural choice is to keep the harvesting core shared
across applications and vary only the application-specific front end.
The shared core consists of the connecting-rod linkage, gear train, DC
motor/generator, rectification and storage stage, and event-triggered
power gate. The waste-bin variant adds an ultrasonic range sensor and
uses long-range LoRa, while the door and cabinet variants omit the
sensor and use short-range LoRa. This factoring allows one hardware
platform to support both subclasses introduced in
Section~\ref{sec:introduction}: \emph{event-only} applications, where
the access event itself is the data, and \emph{event-triggered sensing}
applications, where the access event also powers a measurement.

\subsection{Batteryless Edge Node}
\label{sec:edge-node}

The edge node consists of four functional blocks: mechanical capture and
amplification, electrical generation and storage, event-triggered power
gating, and sensing-and-uplink.

\subsubsection{Mechanical Capture and Amplification}

A compact connecting-rod linkage transfers hinge motion to the
harvesting core. This linkage converts opening and closing motion
into rotation at the gear-train input (Section~\ref{sec:q2_gears}). A three-stage spur-gear train
then amplifies the low hinge speed to the operating range of the
generator. The final design uses a 1:42.6 ratio derived from the
measured hinge-speed range and generator characterization.

\subsubsection{Electrical Generation and Storage}

The amplified rotation drives a 24~V  DC motor used in 
generator mode. Because the generated voltage depends on hinge 
direction and speed, the motor output is rectified on a custom 
PCB. The rectified output charges a 1000~$\mu$F, 25~V capacitor, which 
serves as the energy buffer for one access event. The capacitor feeds a 
buck converter that provides a regulated 3~V rail for the MCU, sensor, 
and LoRa radio. This regulation stage decouples the electronics from the 
time-varying generator output: the capacitor may charge to a higher 
voltage during harvesting, while the active electronics receive a stable 
supply during the wake--sense--transmit cycle. Once the capacitor voltage 
falls below the converter's minimum input voltage ($\approx$3~V), the 
converter turns off; energy remaining below this point is unusable by 
the node. We use a capacitor rather than a rechargeable battery to 
support repeated charge--discharge cycles without scheduled replacement, 
consistent with \textbf{[D1]}. The complete power path, rectifier, storage 
capacitor, event-triggered power gate, and buck-regulated 3~V rail, is 
integrated on the custom PCB.

\subsubsection{Event-Triggered Power Gating}

A switch separates energy harvesting from computation and transmission.
While the hinge is moving, the switch isolates the electronics so that
harvested energy accumulates in the capacitor. Once the hinge reaches
the triggering position, the switch connects the charged buffer to the
regulator and microcontroller. This sequencing prevents the MCU and
radio from drawing power before the buffer has reached a usable voltage. The trigger mechanism is deployment-specific. The bin variant uses a tilt switch actuated by lid closure. The door and cabinet variants use limit switches that can be placed at the fully closed position, fully open position, or both, depending on whether the application needs single or dual-event capture.

\subsubsection{Sensing and Uplink}

The work performed after wake-up depends on the application subclass.
For event-only deployments, such as doors and cabinets, no sensor is
required: the packet carries the node ID and event type, and the hub
assigns the arrival timestamp. For event-triggered sensing, such as
waste-bin monitoring, the MCU triggers an HC-SR04 ultrasonic sensor
mounted inside the lid cavity and includes the fill-level reading in the
uplink packet. Computation runs on a low-power ATmega328P~\cite{ref23} that boots only
after the power gate closes. The radio configuration follows the
deployment context~\textbf{[D4]}. The bin variant uses an RA-02 LoRa
module based on the SX1278~\cite{ref24} at SF10 for outdoor long-range
uplinks. The door and cabinet variants use the same module at SF6 for
short-range indoor links to nearby hubs. SF6 is chosen for indoor use
because the receiver is only meters away, the event-only payload is
small, and the shorter time-on-air reduces transmission energy, leaving
more margin for less predictable user-driven door and cabinet motions.


\section{Harvesting System Design}
\label{sec:harvesting}

The harvesting system is sized for one transaction rather than average power. In motion-coupled sensing, the central requirement is that a single access motion must charge the energy buffer enough to complete one wake--sense--transmit cycle. We size the system for the waste-bin variant because it is the highest-energy workload: each actuation must power ultrasonic sensing and a long-range SF10 LoRa uplink. Door and cabinet deployments are lower-energy event-only workloads, although their reliability depends on mounting geometry and user actuation behavior. The non-obvious challenge is that hinge speed (17.85--27.8\,RPM) is far below the generator required speed, and the harvesting window ($\sim$1.2\,s) is fixed by human behavior, neither parameter can be designed around, only matched. This section develops the sizing procedure. We first measure the energy
required by one transaction, then characterize the hinge motion
available during routine use. We use these measurements to select the
generator, derive the gear ratio, and choose the capacitor based on required energy. We address following research questions in this work.

\begin{enumerate}[leftmargin=*,label=\textbf{Q\arabic*:},itemsep=0.5ex]
\item \textbf{How much energy does one long-range wake--sense--transmit cycle require?}

\item \textbf{Can routine hinge motion, captured with low-cost
      commodity hardware, supply this energy across observed access
      speeds?}
\end{enumerate}

Q1 is answered through direct measurement of the node's energy budget
(Section~\ref{sec:q1}). Q2 is addressed through hinge-motion
characterization (Section~\ref{sec:q2_behavior}) and drivetrain design
(Section~\ref{sec:q2_gears}).

\subsection{Per-Cycle Energy Budget (Q1)}
\label{sec:q1}

\begin{figure}[t]
\centering
\includegraphics[width=1\linewidth]{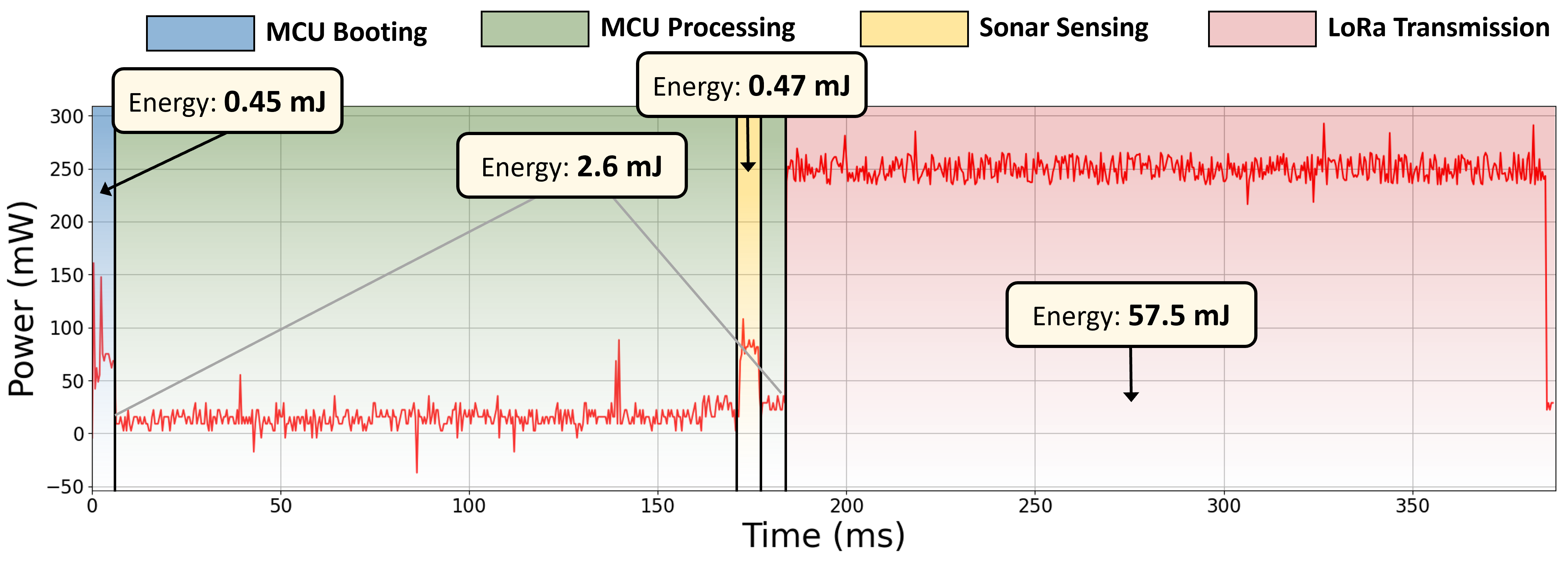}
\caption{Power waveform for single
sensing-and-uplink cycle.}
\label{fig:power_waveform}
\end{figure}

We measured the bin variant in a lab configuration using an Arduino Pro
Mini based on the ATmega328P, an HC-SR04 ultrasonic sensor, and an
RA-02 LoRa module. The measurement captures the full active cycle: MCU
boot, ultrasonic sensing, processing, and uplink. MCU startup consumes
$\approx$0.45~mJ; ultrasonic sensing and processing consume
$\approx$3.0~mJ; and LoRa transmission dominates at
$\approx$57.5~mJ. The total measured energy for one bin transaction is
therefore $\approx$61~mJ. Figure~\ref{fig:power_waveform} shows the
measured power waveform. The door and cabinet variants retain the same RA-02 LoRa radio but use
SF6 instead of SF10, reducing packet airtime and transmission energy
demand~\cite{ref50}. They also omit ultrasonic sensing. A harvester
sized for the 61~mJ bin transaction therefore provides additional energy
margin for these short-range event-only variants, motivating the
asymmetric evaluation strategy used in Section~\ref{sec:evaluation}.

\subsection{Hinge-Motion Characterization (Q2)}
\label{sec:q2_behavior}

\begin{figure}[t]
\centering
\includegraphics[width=0.8\columnwidth]{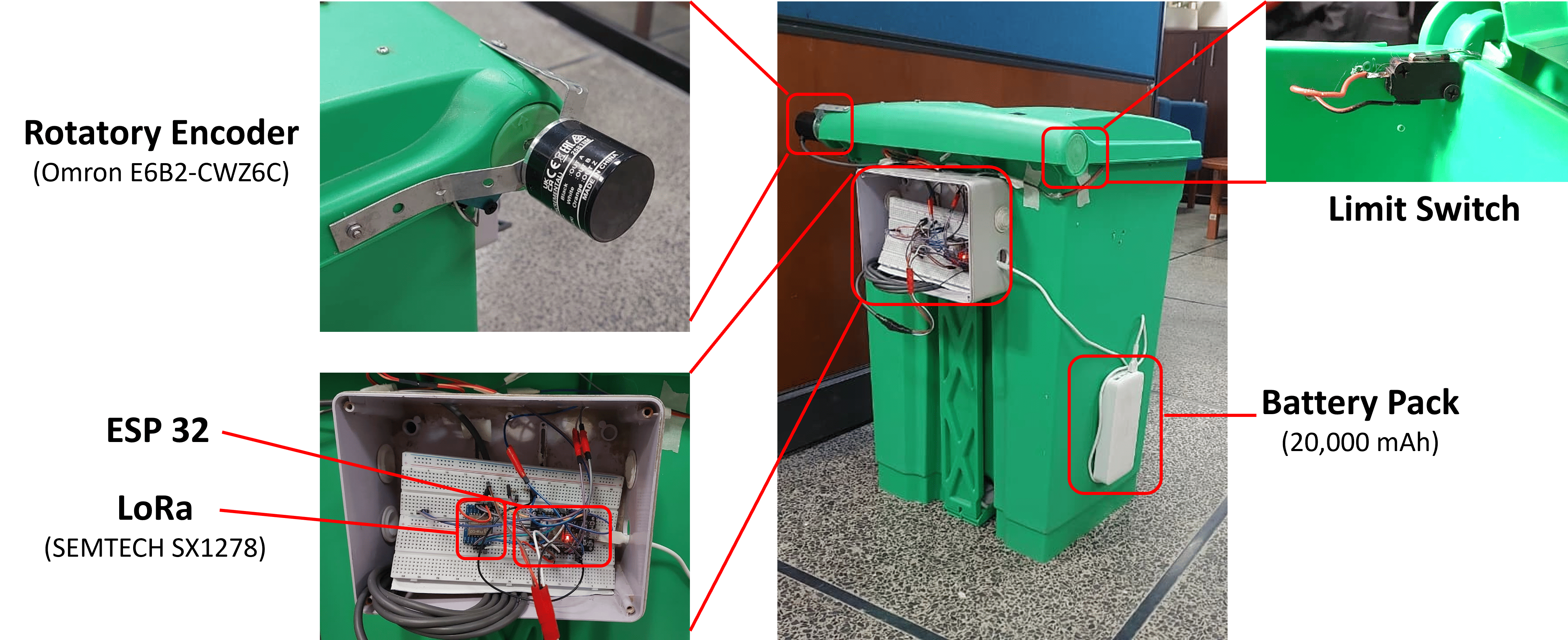}
\caption{Lid-motion instrumentation with encoder, limit switch, and
ESP32 logging at millisecond resolution.}
\label{fig:instrumentation}
\end{figure}

To determine whether routine lid motion can supply the 61~mJ bin
budget, we instrumented a 50-liter hinged-lid bin with an Omron
E6B2-CWZ6C rotary encoder~\cite{ref35} and a limit switch at the
fully closed position (Fig.~\ref{fig:instrumentation}). An ESP32 logged
the encoder and switch signals at millisecond resolution and forwarded
them via LoRaWAN for extraction of opening angle, opening duration, and
closing duration.

\begin{figure}[t]
\centering
\includegraphics[width=\linewidth]{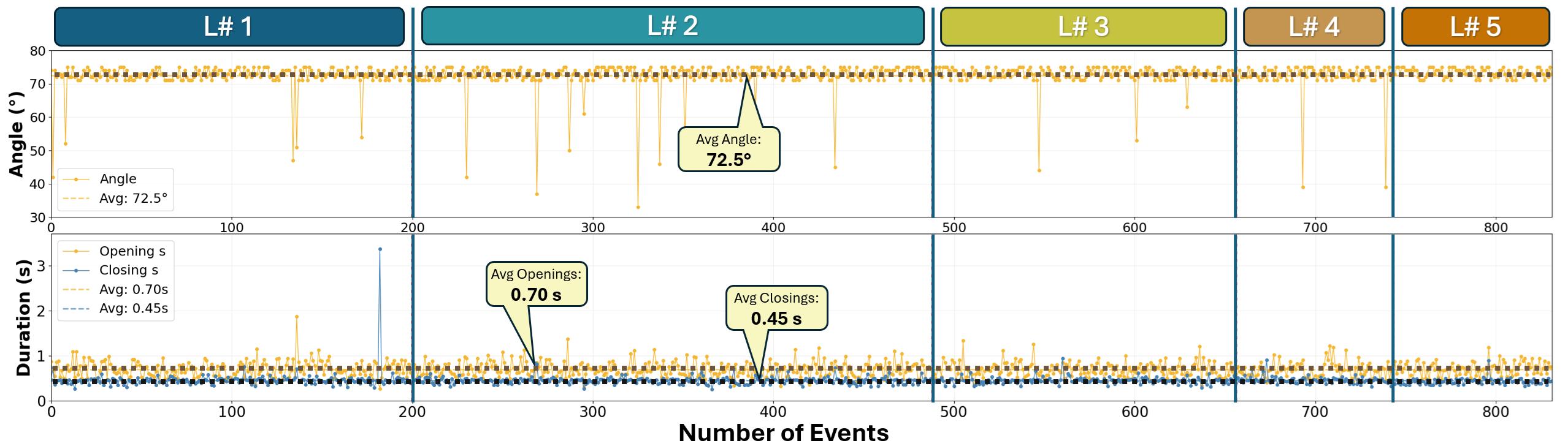}
\caption{Temporal and angular lid characteristics}
\label{fig:opening_angle}
\end{figure}

Across 832 lid interactions at five campus locations (Fig.~\ref{fig:opening_angle}), the mean opening angle was 72.5$^\circ$, the mean opening duration was 0.70~s, and the mean closing duration was 0.45~s. Two findings drive the harvester design. First, closing is gravity-assisted and more repeatable than opening, making it the preferred trigger point for wake-up. Second, the observed hinge speeds, 17.85--27.8~RPM, are far below the operating range of a commodity DC motor used as a generator. The drivetrain must therefore amplify hinge speed before electrical generation.

\subsection{Generator Selection and Gear Train Design (Q2)}
\label{sec:q2_gears}

\subsubsection{Motor-as-Generator Characterization}

\begin{figure}[t]
\centering
\includegraphics[width=0.7\linewidth]{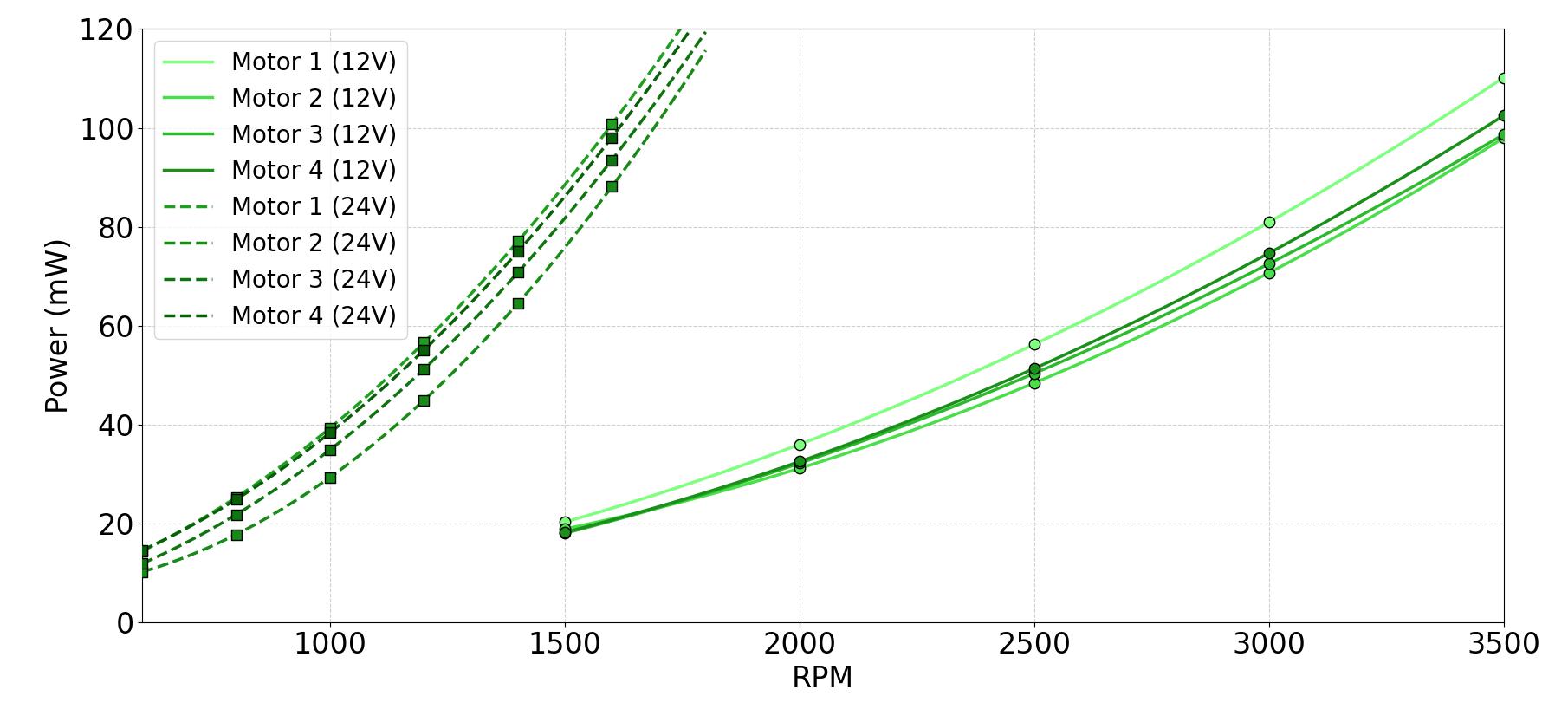}
\caption{Power vs.\ RPM for eight DC motors under
470~$\Omega$ load.}
\label{fig:motor_comparison}
\vspace{-14pt}
\end{figure}

We evaluated eight commodity permanent-magnet DC motors of similar size
($\approx$28~mm diameter, 50~mm length): four rated at 12~V and four
rated at 24~V. Each motor was tested on a variable-speed rig from 600
to 5000~RPM under a 470~$\Omega$ reference load, representing the
average system load (Fig.~\ref{fig:motor_comparison}).

To meet the 61~mJ bin budget, the generator must provide roughly
51~mW over the $\approx$1.2~s harvesting window formed by opening and
closing. The 12~V motors reached this target only above 2200~RPM,
requiring high mechanical amplification. The 24~V motors reached the
same target at approximately 1100~RPM, reducing the required gear ratio
by about half at the same volume price of \$1--1.5 per unit. We
therefore select a 24~V DC motor for the harvester.

\subsubsection{Gear Ratio Design}
The required gear ratio follows from the measured hinge-speed range and
the generator target. Mapping 17.85--27.8~RPM at the hinge to roughly
1100~RPM at the generator requires an amplification of about
42.6$\times$. We implement this using a three-stage spur-gear train
assembled from commodity parts: a 13-tooth motor pinion, two compound
gears with 38-tooth outer and 13-tooth inner gears, and a 65-tooth
driven gear:

{\footnotesize
\[
13T \rightarrow 38T/13T \rightarrow 38T/13T \rightarrow 65T .
\]
}

This train yields a 1:42.6 ratio, mapping the observed hinge-speed band
to 760--1184~RPM at the generator. The lower end of this range is below
the 1100~RPM nominal target, so slow or partial actuations may not
harvest enough energy for a complete transaction. In normal use,
however, capacitor buffering and trigger-on-close sequencing allow full
lid actuations to accumulate sufficient energy before wake-up. The field
results in Section~\ref{sec:eval_bin} show that most full actuations
cross the operating threshold, while the door and cabinet evaluations in
Section~\ref{sec:evaluation} show that the same ratio provides
sufficient energy for lower-budget event-only workloads.

\subsection{Capacitor Sizing and Wake-Up Sequencing}
\label{sec:harvester}

\begin{figure}[t]
\centering
\includegraphics[width=0.6\linewidth]{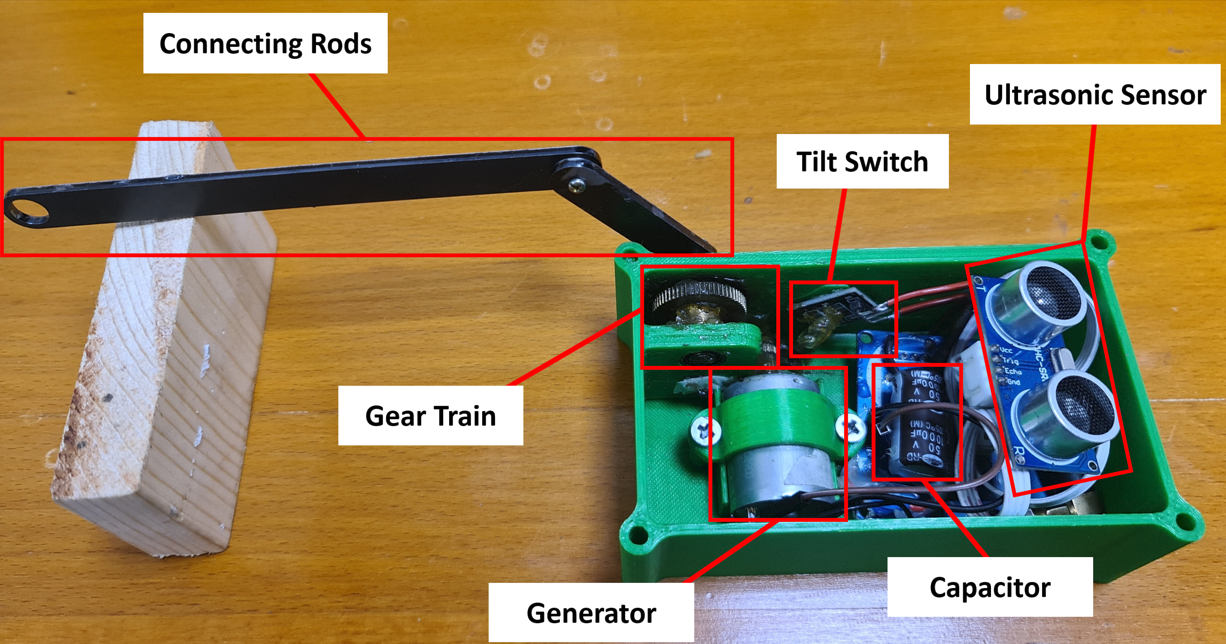}
\caption{Implementation of the harvesting module}
\label{fig:harvester_module}
\vspace{-10pt}
\end{figure}

\begin{figure*}[t]
\centering
\begin{subfigure}[t]{0.295\textwidth}
    \centering
    \includegraphics[width=\linewidth]{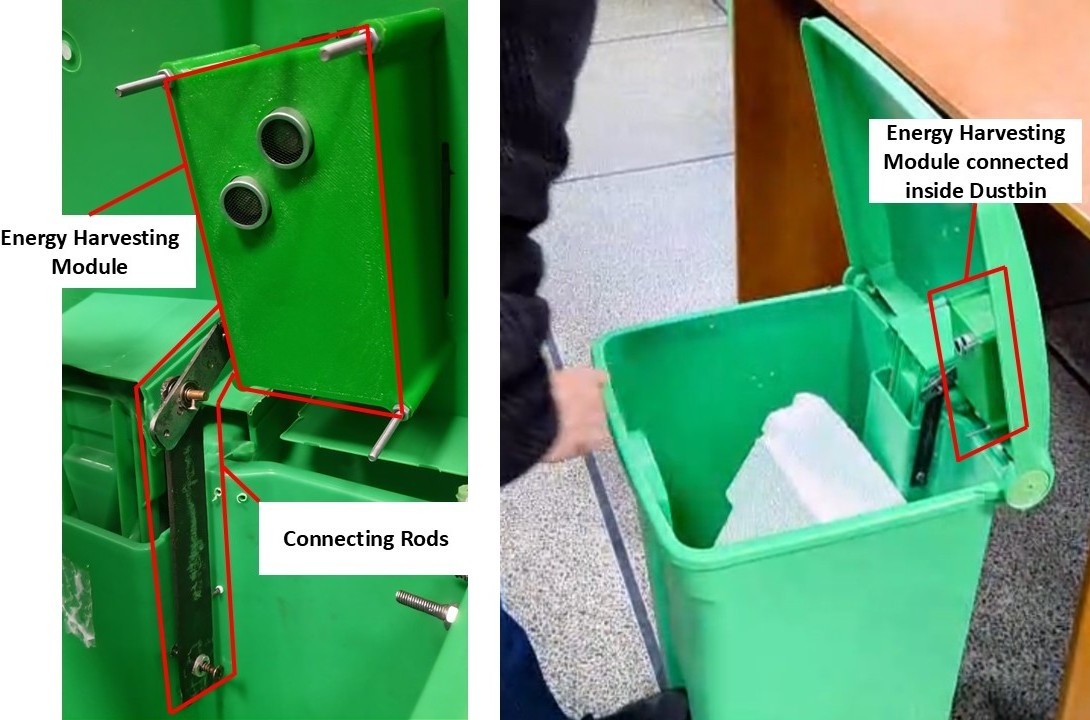}
    \caption{Waste bin.}
    \label{fig:bin_module}
\end{subfigure}
\hfill
\begin{subfigure}[t]{0.26\textwidth}
    \centering
    \includegraphics[width=\linewidth]{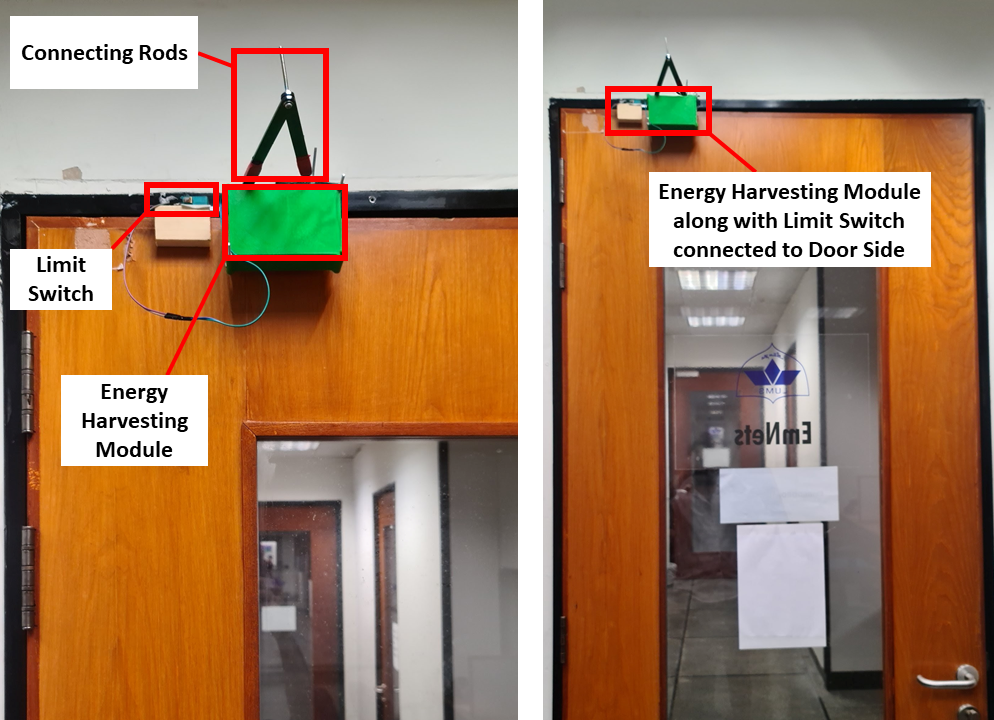}
    \caption{Room door.}
    \label{fig:door_module}
\end{subfigure}
\hfill
\begin{subfigure}[t]{0.26\textwidth}
    \centering
    \includegraphics[width=\linewidth]{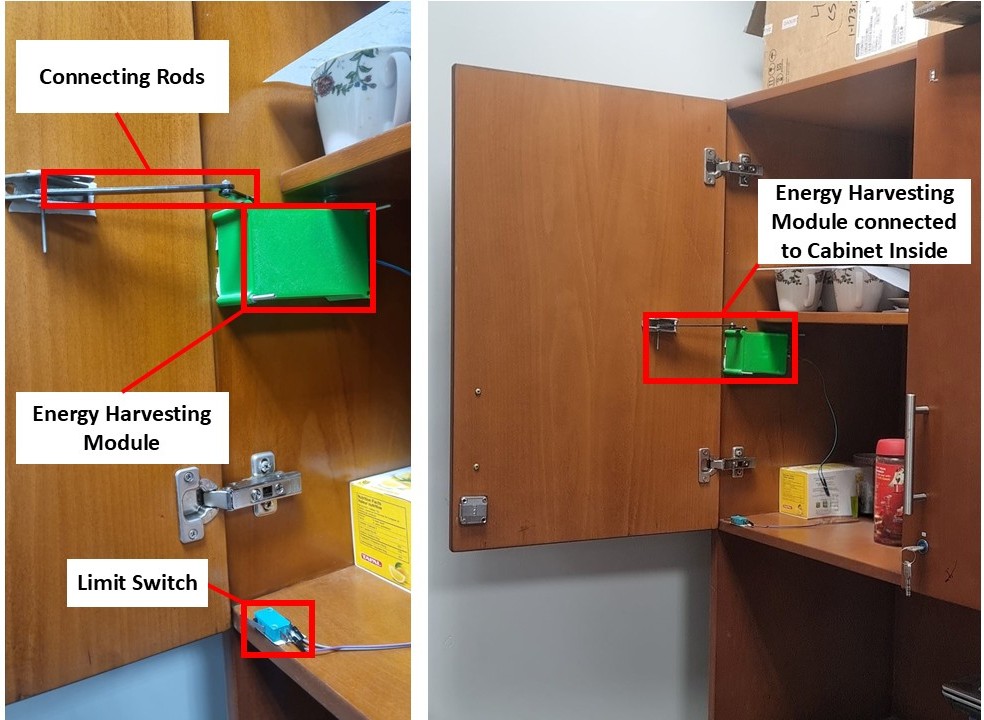}
    \caption{Office cabinet.}
    \label{fig:cabinet_module}
\end{subfigure}
\caption{Three deployments sharing the same harvesting core and RA-02 LoRa module.}
\label{fig:instantiations}
\end{figure*}

Figure~\ref{fig:harvester_module} shows the implemented mechanical
assembly. The remaining design parameters are the storage capacitance
and the wake-up threshold. We use a 1000~$\mu$F, 25~V capacitor as the local energy buffer. The motor output is rectified on a custom PCB and used to charge the capacitor. A buck converter then regulates the capacitor voltage to a stable 3~V rail for the MCU, ultrasonic sensor, and LoRa radio. The converter operates only while its input voltage remains above its minimum. Once the capacitor falls below $\approx$3~V, the converter turns off, so energy stored below this voltage is unusable.

The usable energy in the capacitor is therefore the energy between the
charged voltage and the converter cutoff:
{\footnotesize
\[
E_{\mathrm{usable}} =
\frac{1}{2}C\left(V_{\mathrm{charged}}^2 -
V_{\mathrm{cutoff}}^2\right).
\]
}

With $C=1000~\mu$F, $V_{\mathrm{cutoff}}\approx3$~V, and
$E_{\mathrm{usable}}\approx61$~mJ, the required charged voltage is
{\footnotesize
\[
V_{\mathrm{charged}} =
\sqrt{\frac{2E_{\mathrm{usable}}}{C} + V_{\mathrm{cutoff}}^2}
=
\sqrt{\frac{2(61\times10^{-3})}{1000\times10^{-6}} + 3^2}
\approx 11.45~\mathrm{V}.
\]
}

We therefore use 11.5~V as the empirical minimum operating threshold
for the bin variant. At this voltage, the 1000~$\mu$F capacitor stores
approximately 61.6~mJ of usable energy above the 3~V cutoff:

{\footnotesize
\[
\frac{1}{2}(1000\times10^{-6})(11.5^2 - 3^2)
\approx 61.6~\mathrm{mJ}.
\]
}

The 25~V capacitor rating provides voltage headroom, while the selected
capacitance is large enough to support LoRa peak current without
excessive rail droop and small enough to recharge within one access
event. Wake-up sequencing uses the closing asymmetry observed in
Section~\ref{sec:q2_behavior}. Because closing is gravity-assisted and
more consistent across users, the bin variant triggers on lid closure.
The tilt switch for bins, or limit switch for doors and cabinets,
connects the charged capacitor to the regulator only after the hinge
reaches its triggering position. This prevents the MCU and radio from
drawing energy during harvesting. Once powered, the electronics complete
the active transaction in under 0.5~s and then return to zero idle current.

\section{Platform Variants and Cost}
\label{sec:variants}

We instantiate the platform in three hinged-access deployments using
two electronics configurations. All variants share the same harvesting
core and power path: mechanical assembly, rectifier, capacitor buffer,
event-triggered power gate, buck-regulated 3~V supply, microcontroller,
and RA-02 LoRa radio, as described in
Sections~\ref{sec:design} and~\ref{sec:harvesting}. The variants differ
only in sensing workload and radio configuration. The waste-bin variant
adds ultrasonic sensing and uses long-range LoRa for outdoor uplinks;
the door and cabinet variants omit sensing and use short-range LoRa to
nearby indoor hubs.

\subsection{Waste Bin Lid (Event-Triggered Sensing)}
\label{sec:variant-bin}

The waste-bin module attaches inside the lid cavity of a standard 50-liter hinged-lid bin (Fig.~\ref{fig:bin_module}). An HC-SR04 ultrasonic sensor~\cite{ref25} faces downward from the top, maintaining acoustic line-of-sight to the waste surface while the enclosure protects the electronics and drivetrain from rain and debris. When the lid closes, a tilt switch connects the charged capacitor to the 3~V rail. The MCU triggers the sensor and transmits a long-range LoRa packet containing the node ID and fill-level reading. This variant targets outdoor campus and municipal deployments where the gateway may be hundreds of meters away.

\subsection{Room Door (Event-Only)}
\label{sec:variant-door}

The door variant mounts the same harvesting core on the door frame near
the hinge (Fig.~\ref{fig:door_module}). Because the access
event itself is the information of interest, no additional sensor is
required. A limit switch at the fully closed rest position powers the
node on door close, triggering a short-range LoRa packet containing the
node ID and event type. The in-room hub assigns the arrival timestamp,
so the batteryless node does not need to maintain a clock between
events. Target applications include occupancy monitoring,
energy-management triggers, and access logging in office or residential
settings.

\subsection{Office/Home Cabinet (Event-Only)}
\label{sec:variant-cabinet}

The cabinet variant uses the same event-only configuration as the door
variant, adapted to a smaller hinge geometry
(Fig.~\ref{fig:cabinet_module}). A limit switch powers the node when the
cabinet reaches the configured trigger position, and the node transmits
a short-range LoRa packet containing the node ID and event type. As in
the door deployment, the in-room hub timestamps packet arrival. The
field deployment in Section~\ref{sec:evaluation} shows that the
shared harvesting core provides sufficient energy for this lower-power
event-only workload.

\fakepara{Dual-event capture}
The cabinet variant can also support dwell-time logging by placing
limit switches at both the fully open and fully closed positions. In
this configuration, the opening motion powers and triggers an
``opened'' packet, while the closing motion powers and triggers a
``closed'' packet. The hub estimates dwell time from the arrival times
of the two packets. This option adds one more switch and one more packet per access cycle, each powered by its corresponding hinge motion. Target applications include pharmaceutical compliance monitoring, office asset tracking, and smart-home appliance logging.

\subsection{Cost Analysis}

\begin{table}[t]
\centering
\caption{BOM per Unit (USD, Volume $\geq$1,000 units)}
\label{tab:cost}
\adjustbox{max width=0.85\columnwidth}{%
\begin{tabular}{@{}lcc@{}}
\toprule
\textbf{Component} & \textbf{Long-Range Bin} & \textbf{Short-Range Door/Cabinet} \\
\midrule
MCU (ATmega328P)              & 1.50 & 1.50 \\
LoRa Module (SX1278)          & 3.80 & 3.80 \\
Ultrasonic Sensor (HC-SR04)   & 1.00 & --   \\
DC Motor/Generator (24~V)     & 1.50 & 1.50 \\
Capacitor (1000~$\mu$F, 25~V) & 0.50 & 0.50 \\
Power Management PCB          & 3.50 & 3.50 \\
Gear Train (brass/plastic)    & 3.00 & 3.00 \\
Connecting Rods               & 2.00 & 2.00 \\
Enclosure (ABS plastic)       & 3.00 & 3.00 \\
Limit/Tilt Switch             & 0.50 & 0.50 \\
\midrule
\textbf{Total BOM Cost} & \textbf{\$20.30} & \textbf{\$19.30} \\
\bottomrule
\end{tabular}}
\vspace{-10pt}
\end{table}

Table~\ref{tab:cost} reports the per-unit BOM at volume pricing. The
long-range bin variant costs \$20.30, while the short-range
door/cabinet variant costs \$19.30. The \$1.00 difference is due only
to the ultrasonic sensor; the harvesting core, MCU, radio, power
management PCB, enclosure, and mechanical assembly are shared across
all deployments. As a result, the platform supports multiple
applications with a common supply chain and manufacturing process, while
the BOM changes by only $\approx$5\% between sensing and event-only
configurations.

\section{Evaluation}
\label{sec:evaluation}

Our evaluation follows the platform's energy hierarchy. It does not ask
whether a harvester can generate energy in isolation; it asks whether
naturally occurring state-changing motions can complete one bounded sensing
transaction with high per-event reliability. We first field-validate the
highest-energy variant, waste-bin fill-level sensing, where each lid
actuation must power both ultrasonic measurement and a long-range LoRa
uplink. We then field-deploy the same harvesting core for door and cabinet
event-only workloads at deployment scale (1{,}870 and 1{,}636 actuations respectively), testing
whether the bin-sized energy envelope transfers across hinge geometries
without hardware redesign. The three deployments together provide per-event
reliability evidence across the full range of motion-coupled sensing
workloads, from outdoor sensing-and-uplink to indoor event-only reporting.

\subsection{Waste Bin: Full Field Validation}
\label{sec:eval_bin}

\subsubsection{Multi-Location Deployment}

\begin{figure}[t]
\centering
\includegraphics[width=0.46\textwidth]{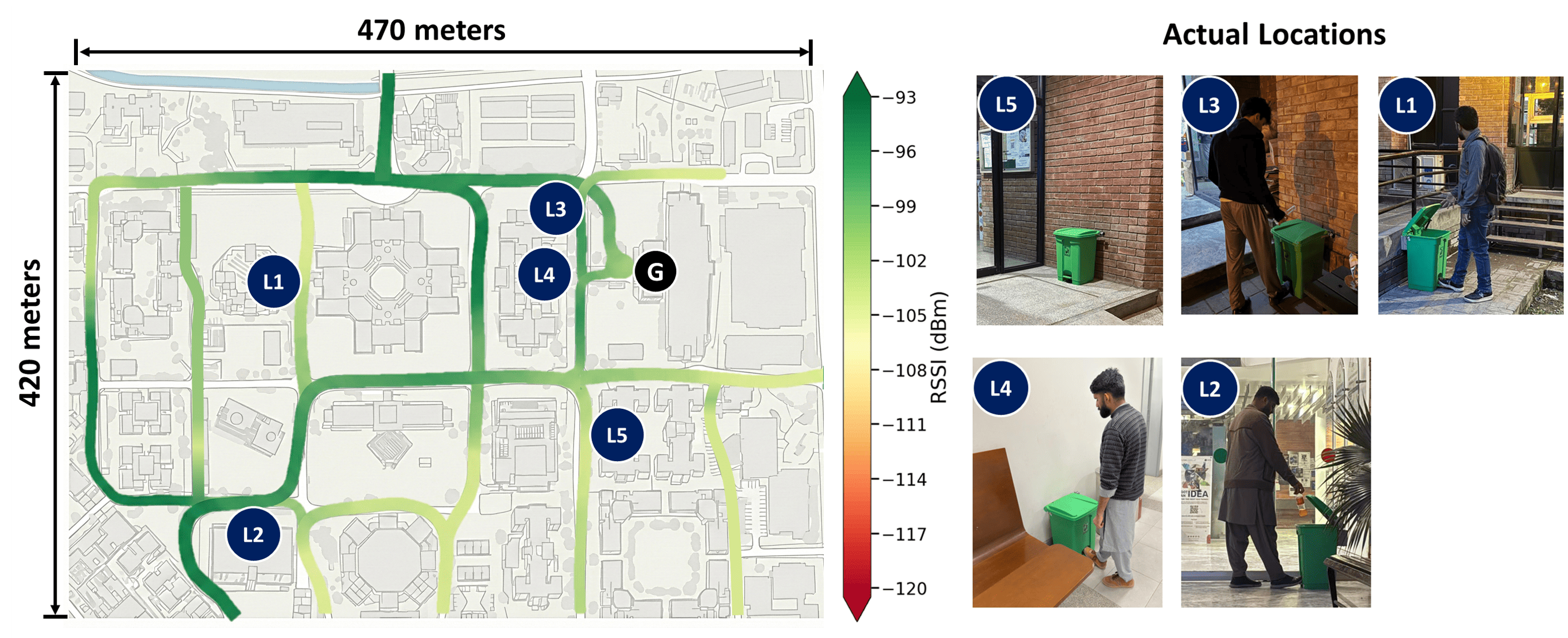}
\caption{Geographic distribution of bin deployment locations.}
\label{fig:deployment_map}
\vspace{-10pt}
\end{figure}

We sequentially deployed the bin module at five moderate- to
high-traffic campus locations (Fig.~\ref{fig:deployment_map}) under
real outdoor conditions: 18--38\textdegree{}C temperature,
40--85\% humidity, and occasional rainfall. The ultrasonic transducer
faced downward inside the lid cavity, while the enclosure protected electronics and drivetrain from rain and debris~\cite{ref6,ref44}. A rooftop LoRaWAN
gateway received packets containing timestamp, bin ID, fill-level
reading, and RSSI.

\fakepara{Deployment sites}
\begin{itemize}[nosep, leftmargin=1.8em, itemsep=1.5pt]
  \item[\scalebox{0.75}{\circG{}}]          \textbf{Gateway}: campus rooftop.
  \item[\scalebox{0.75}{\circnumblue{L1}}]  \textbf{Library}: consistent academic-hour traffic.
  \item[\scalebox{0.75}{\circnumblue{L2}}]  \textbf{Business School}: class-transition peaks.
  \item[\scalebox{0.75}{\circnumblue{L3}}]  \textbf{Cafe Entrance}: meal-time rapid successive actuations.
  \item[\scalebox{0.75}{\circnumblue{L4}}]  \textbf{Cafeteria}: traffic concentrated around meals.
  \item[\scalebox{0.75}{\circnumblue{L5}}]  \textbf{Dormitories}: Variable traffic with morning, evening peaks.
\end{itemize}

Because the system is event-driven, reliability is measured per access:
one lid actuation should produce one complete sensing-and-uplink cycle.
No periodic polling occurs between accesses; for a lidded bin, the last
reported fill level remains valid until the next lid-opening event.

\begin{figure*}[t]
\centering
\includegraphics[width=0.75\textwidth]{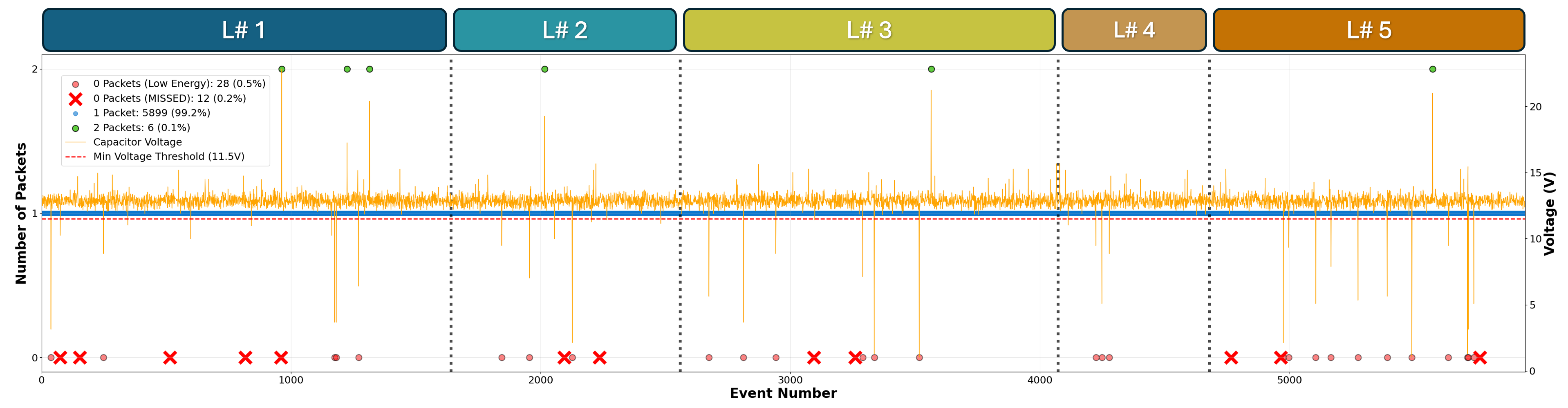}
\caption{Packet transmissions and capacitor voltage over 5,945 bin
actuations.}
\label{fig:packet_voltage}
\end{figure*}

\fakepara{Aggregate results}
Across all sites, the platform recorded \textbf{5,945 lid actuations}
and \textbf{5,905 successful transmissions} over three weeks,
achieving \textbf{99.3\% per-event reliability}
(Table~\ref{tab:deployment}). Figure~\ref{fig:packet_voltage} shows packet outcomes and capacitor
voltage over the deployment. The red dashed line marks the 11.5~V
minimum transmission threshold. Most actuations produced exactly one
packet; a few rapid re-openings produced two packets before the
capacitor fully discharged.

\begin{table}[t]
\centering
\caption{Per-location bin deployment results.}
\label{tab:deployment}
\scriptsize
\resizebox{0.85\columnwidth}{!}{%
\begin{tabular}{@{}lccc@{}}
\toprule
\textbf{Location} & \textbf{Actuations} & \textbf{Packets Sent} &
\textbf{Success (\%)} \\
\midrule
L1 Library         & 1,640 & 1,630 & 99.3 \\
L2 Business School &   918 &   913 & 99.4 \\
L3 Cafe Entrance   & 1,514 & 1,506 & 99.4 \\
L4 Cafeteria       &   607 &   604 & 99.5 \\
L5 Dormitories     & 1,266 & 1,252 & 98.8 \\
\midrule
\textbf{Total/Avg} & \textbf{5,945} & \textbf{5,905} & \textbf{99.3} \\
\bottomrule
\end{tabular}}
\vspace{-10pt}
\end{table}

\fakepara{Packet distribution}
Of 5,945 lid actuations, 5,899 events (99.3\%) produced exactly one
packet. Six events (0.1\%) produced two packets after rapid re-opening.
Forty events produced no packet: 28 partial actuations (0.5\%) did not
charge the capacitor above 11.5~V, and 12 events (0.2\%) were
attributed to LoRaWAN channel loss. Post-deployment inspection showed
no water ingress, minimal dust accumulation, and no visible corrosion
or mechanical degradation across the five locations~\dg{1}\dg{2}.

\subsubsection{Ultrasonic Fill-Level Accuracy}

\begin{figure}[!htbp]
\centering
\includegraphics[width=0.8\columnwidth]{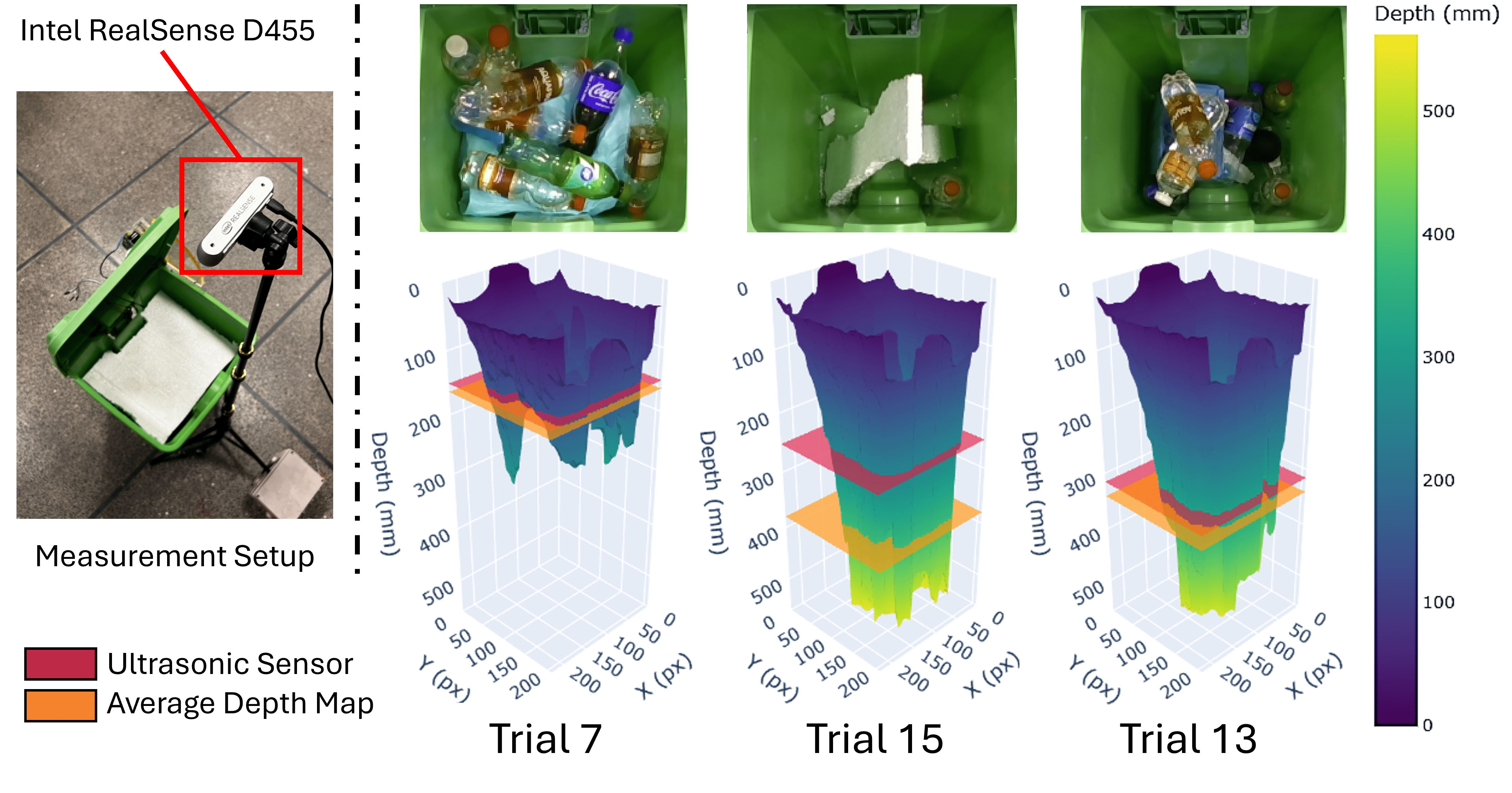}
\caption{Comparison of depth-camera and ultrasonic
fill-level.}
\label{fig:depth_validation}
\vspace{-10pt}
\end{figure}

Transmission reliability shows that one access event can power sensing
and communication; the bin variant must also report a useful fill-level
estimate. We therefore evaluate ultrasonic accuracy against
depth-camera ground truth to separate sensing error from communication
loss.

We mounted an Intel RealSense depth camera above a retrofitted bin and recorded 50 trials across varying waste configurations (Fig.~\ref{fig:depth_validation}), spanning the full bin depth range to expose any distance-dependent spread in ultrasonic readings. For each trial the walls were masked from the raw depth map to isolate the bin interior, and the masked values spatially averaged to produce a ground-truth fill level estimate. To mitigate setup misalignments, ultrasonic readings were adjusted for linear scaling and bias offset prior to comparison against the ground truth.

\begin{figure}[!htbp]
\centering
\begin{minipage}[c]{0.32\columnwidth}
  \centering
  \footnotesize
  \setlength{\tabcolsep}{3pt}
  \renewcommand{\arraystretch}{1.15}
  \begin{tabular}{@{}l r@{}}
    \toprule
    \textbf{Metric} & \textbf{Value} \\
    \midrule
    Trials         & 50 \\
    MAE (mm)       & 19.08 \\
    Min error (mm) & 0.3 \\
    Max error (mm) & 63.9 \\
    Std dev (mm)   & 15.9 \\
    \bottomrule
  \end{tabular}
  \\[2pt]
  {\footnotesize(a) Statistical summary}
  \label{tab:depth_stats}
\end{minipage}%
\hfill%
\begin{minipage}[c]{0.64\columnwidth}
  \centering
  \includegraphics[width=\linewidth]{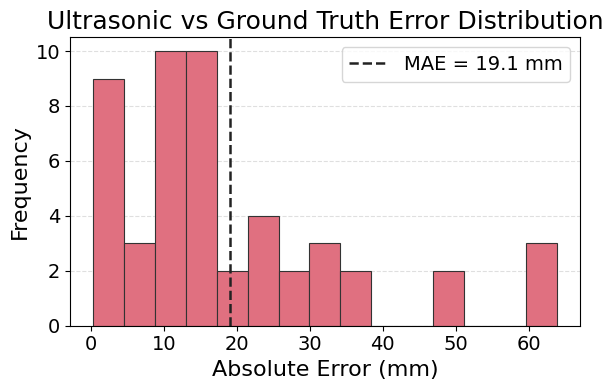}
  \vspace{1pt}
  {\footnotesize(b) Absolute error distribution}
\end{minipage}
\caption{Ultrasonic fill-level accuracy: statistical summary (a) and absolute error distribution across 50 trials (b).}
\label{fig:depth_stats_and_hist}
\end{figure}

The HC-SR04's $\approx$15$^\circ$ acoustic cone reports the dominant
reflection within its sensing region rather than a point measurement.
It therefore behaves as a coarse surface estimator: accurate when waste
is distributed within the cone, but biased by protruding objects
directly beneath the sensor. Across 50 trials, the system achieves a
mean absolute error of 19.08\,mm
(Fig.~\ref{fig:depth_stats_and_hist}), sufficient for coarse bulk
fill-level estimation while identifying protrusions as the dominant
failure mode.

\noindent\textbf{Operational interpretation.}
Collection scheduling typically acts on coarse fill categories, such as
\emph{empty}, \emph{quarter}, \emph{half}, \emph{three-quarter}, and
\emph{full}, rather than on millimeter-level fill curves~\cite{ref8}.
For the bin's 500~mm usable depth, these five categories correspond to
roughly 100~mm per bucket, so an MAE meaningfully below this threshold
suffices for collection routing. The error tail is dominated by a known
geometric failure mode, a tall protrusion within the acoustic cone, which can be addressed through firmware or packaging refinements,
such as multi-ping median filtering or a small acoustic baffle, and is
orthogonal to the paper's energy-harvesting contribution.

\subsubsection{Communication Range}
LoRa SF10 range was characterized at the deployment site (125\,kHz BW,
CR\,4/8, 20\,dBm, 8-symbol preamble, CRC); all five bin locations
operated within the SX1278 link budget~\cite{ref24,ref28}, with RSSI
degraded by foliage and building obstruction as expected.

\subsubsection{Mechanical Stress Test}

\begin{figure}[t]
\centering
\includegraphics[width=0.9\columnwidth]{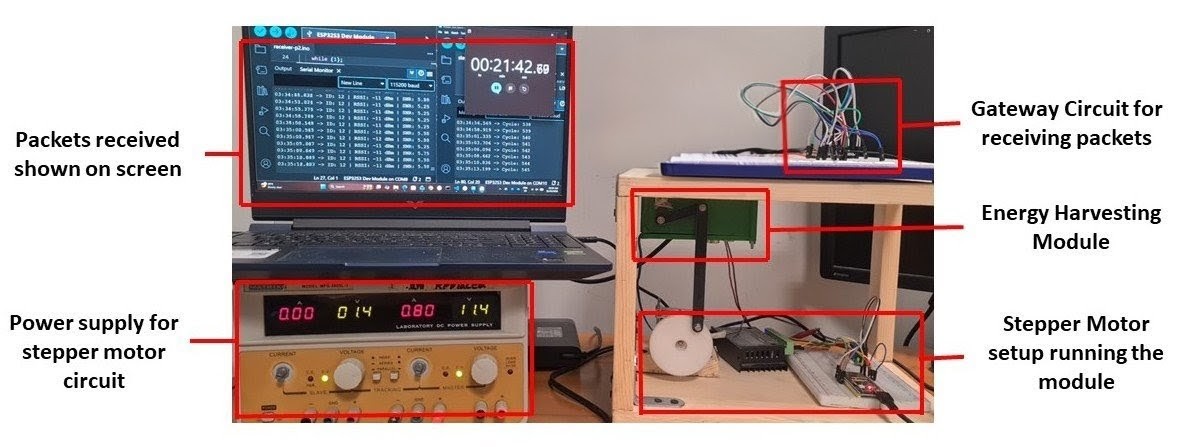}
\caption{Mechanical stress test setup}
\label{fig:stress_test}
\vspace{-10pt}
\end{figure}

We subjected the mechanical assembly to a 10,000-iteration accelerated
life test using the stepper-motor-operated actuator in
Figure~\ref{fig:stress_test}. The module transmitted successfully after
every actuation. Post-test inspection showed no visible gear tooth
wear, connecting-rod fatigue, mechanical
misalignment, or enclosure damage. The full
7 hour test was recorded with visible timestamps
throughout~\cite{StressTestVideo2025}.

\subsection{Room Door: Field Deployment}
\label{sec:eval_door}

The connecting-rod module was mounted on a standard interior door hinge
in a university office without modifying the door or frame. A
short-range LoRa receiver was placed in the same room. We conducted
\textbf{1{,}870 door actuations} under normal use conditions documented in the supplementary
video~\cite{DoorEvalVideo2026}.

\textbf{Result:} Table~\ref{tab:door_cabinet_summary} indicates 1{,}724 of 1{,}870 actuations produced a packet,
yielding a \textbf{92\% success rate}. Each packet contained node
ID and event type; the hub assigned the arrival timestamp. Post-deployment
inspection showed no wear or misalignment, and the module added
negligible resistance to door movement~\dg{2}. At deployment scale,
this confirms that the bin-sized energy envelope generalizes to a different 
hinge geometry without redesign, the same harvesting core powers reliable 
event-only reporting on a standard interior door with no battery, no ambient 
harvester, and no hardware modification.

\subsection{Office/Home Cabinet: Field Deployment}
\label{sec:eval_cabinet}

The cabinet trial used the same event-only hardware as the door variant. We conducted \textbf{1{,}636 cabinet actuations} under normal use conditions, documented in the supplementary video~\cite{CabinetEvalVideo2026}.

\textbf{Result:} Table~\ref{tab:door_cabinet_summary} indicates 1{,}535 of 1{,}636 actuations produced a
packet, yielding a \textbf{94\% success rate}. Packet contents
and receiver configuration matched the door deployment. Post-deployment
inspection showed no mechanical wear or alignment issues~\dg{2}. This
second deployment-scale evaluation confirms the energy budget transfers
across a smaller hinge geometry: the same harvesting core powers
reliable event-only reporting on a standard office cabinet without any
hardware redesign relative to the bin or door variants.

\begin{table}[t]
\centering
\caption{Door and cabinet event-only field deployments.}
\label{tab:door_cabinet_summary}
\scriptsize
\setlength{\tabcolsep}{3.2pt}
\begin{tabular}{@{}lcccc@{}}
\toprule
\textbf{Deployment} & \textbf{Actuations} & \textbf{Packets} &
\textbf{Success} & \textbf{Dominant failure mode} \\
\midrule
Room door & 1{,}870 & 1{,}724 & 92\% & weak/partial closure \\
Cabinet & 1{,}636 & 1{,}535 & 94\% & variable user behaviour \\
\bottomrule
\end{tabular}
\vspace{-10pt}
\end{table}

\section{Discussion}
\label{sec:discussion}

\subsection{Scope of Motion-Coupled Sensing}

Motion-coupled sensing targets IoT workloads where state changes are
sparse, mechanically induced, and coupled to motion that can be
harvested. It is therefore not a general replacement for
battery-powered IoT, but a design pattern for access-gated sensing
tasks. In this setting, the relevant reliability metric is not periodic
uptime or battery lifetime, but the probability that one access event
produces one complete sensing transaction. This motivates our
per-event reliability analysis across $\sim$9{,}500 actuations spanning
three access types: bins, doors, and cabinets. The waste-bin variant
defines the platform's energy envelope because it combines ultrasonic
sensing with long-range SF10 LoRa transmission, requiring approximately
61~mJ per cycle. Door and cabinet variants are lower-energy event-only
workloads using short-range SF6 LoRa; their deployments confirm that
the bin-sized energy envelope transfers across hinge geometries at
scale. Motion-coupled sensing is not generic kinetic harvesting: it uses kinetic
energy only when the motion is causally tied to the state being monitored.
It is also distinct from event-driven sensing that assumes a separate power
source and from intermittent computing that assumes scheduled energy arrival
from external sources. Kinetic switches occupy the event-only endpoint;
the harder case evaluated here is event-triggered sensing, where the same
physical access both bounds the validity of the state and supplies the
energy to measure and report that state.






\subsection{Limitations}

The current prototype is limited to hinged access objects with
sufficient angular displacement and a usable fixed anchor point for the
connecting rod. Very small-displacement hinges, recessed hinges,
sliding mechanisms, and objects without suitable mounting points are
outside the present scope. While the three deployments establish
per-event reliability at scale, longer studies are needed to evaluate weather exposure, and mechanical wear over months to years. Finally, the fill-level accuracy study validates coarse bulk
estimation for one bin geometry; additional testing across bin shapes
and waste profiles is needed before making application-specific
accuracy claims.

\section{Conclusion}
\label{sec:conclusion}

We introduced \emph{motion-coupled sensing}, a design pattern for
mechanically gated IoT workloads in which the access event creates the
information, triggers the transaction, and supplies the energy needed
to report it. We demonstrated this principle with an open-source
batteryless platform built around a connecting-rod electromagnetic
harvester, 1:42.6 gear train, 1000~$\mu$F capacitor buffer, and
event-triggered power gate. Sized for the highest-energy workload, the platform powers one
wake--sense--transmit transaction per hinge actuation without scheduled
battery maintenance. In outdoor waste-bin deployments, it achieves
99.3\% per-event reliability over 5{,}945 lid actuations across five
campus locations. Field deployments on indoor doors and cabinets
extend this to lower-energy event-only workloads, achieving
92\% success over 1{,}870 door actuations and 94\%
success over 1{,}636 cabinet actuations using the same harvesting core
without redesign. These results show that mechanically gated sensing tasks can be treated
as self-powered access transactions rather than periodically polled
battery-powered workloads. For this class of IoT deployments, the
motion that changes the state can also serve as the trigger and energy
source for sensing and communication.

\balance
\bibliographystyle{IEEEtran}
\bibliography{references}

\end{document}